\documentclass[twocolumn,aps,prl,superscriptaddress]{revtex4}
\usepackage{graphicx}
\usepackage{bm}
\usepackage{multirow}

\begin{document}

\title{Upper critical field of \emph{p}-wave superconductors with orthorhombic symmetry}


\author{Christopher L\"{o}rscher }
\author{Richard A Klemm}

\affiliation{Department of Physics, University of Central Florida, Orlando, FL 32816-2385 USA}
\author{Jingchuan Zhang}
\author{Qiang Gu}
\affiliation{Department of Physics, University of Science and Technology Beijing, Beijing 100083, China}


\date{\today}

\begin{abstract}
We extend the Scharnberg-Klemm theory of $H_{c2}$ in \textit{p}-wave superconductors with broken symmetry to cases of partially broken symmetry in an orthorhombic crystal, as is appropriate for the more exotic ferromagnetic superconductor UCoGe in strong applied magnetic fields. For some partially broken symmetry cases, $H_{c2}$ can mimic upward curvature in all three crystal axis directions, and reasonable good fits to some of the low-field UCoGe data are obtained.
\end{abstract}


\maketitle

\section{Inroduction} There has long been an interest in the possibility of superconductivity with the paired electrons having an order parameter consisting of a triplet spin configuration and the corresponding odd orbital symmetry [1, 2, 3, 4, 5, 6, 7, 8]. The simplest odd orbital symmetry has the \emph{p}-wave form. In a crystal with a non-cubic structure, there can be a variety of different \textit{p}-wave states [1, 2, 3, 4, 5]. Depending upon the temperature \emph{T}, magnetic field \textbf{\emph{H}}, and pressure \emph{P}, there can be phases corresponding to different triplet states [6, 7, 8]. One of the easiest ways to characterize the \emph{p}-wave states is by measurements of the \emph{T} dependence of the upper critical field $H_{c2}$ [1, 2]. However, when multiple phases are present in the same crystal, as in UPt$_{3}$, a proper analysis requires a variety of experimental results [6, 7].

    Recently, a new class of ferromagnetic superconductors has been of great interest. Presently this class consists of UGe$_{2}$[9], UIr [10], URhGe [11], UCoGe [12], which except for UIr have orthorhombic crystal structures. For URhGe, the superconductivity arises within the ferromagnetic phase. That is also true for UCoGe at ambient pressure, but when sufficient pressure is applied, the ferromagnetic phase disappears, leaving the superconducting phase without any obvious additional ferromagnetism [13, 14]. In the cases of UGe$_{2}$ and UIr, applying pressure within the ferromagnetic phase induces the superconductivity [9, 10]. In addition, polarized neutron experiments have been interpreted as providing evidence for a field induced ferrimagnetic state in UCoGe, with local moments of different magnitudes in opposite directions on the U and Co sites [15]. For a ferromagnetic superconductor with orthorhombic symmetry, the possible order parameter symmetries were given by Mineev [16].

    Hardy and Huxley measured ${\bm B}_{c2}(T)$ of URhGe at ambient pressure in all three crystal axis directions [17]. Using only the slope at $T_{c}$ in each crystal direction as a fitting parameter, they found that the Scharnberg-Klemm theory fit their data quantitatively [17], assuming the crystal-aligned polar state with completely broken symmetry (CBS) [2]. For this state, ${\bm B}_{c2}(T)$ has a distinctly different $T$ dependence for ${\bm B}||\hat{\bm c}$, along the polar axis, than for ${\bm B}||\hat{\bm a}$, and ${\bm B}||\hat{\bm b}$, in the axial plane.  By fitting only the slopes at $T_c$ in these three crystal axis directions, which eliminated the effective mass anisotropy for the closed orbits of the paired electrons in the three crystal directions, as in Landau diamagnetism and in the de Haas-van Alphen effect,  the results implied that the $p$-wave state $|\psi\rangle\propto p_c$ for all applied field directions. This remarkable fit for the low field regime of the superconducting state in URhGe did not require the inclusion of the ferromagnetism into the theory, as the only apparent effect of the magnetization was to give rise to a demagnetization jump in $H_{c2}$ at the superconducting transition temperature, $T_{c}$, beyond which the slopes of $H_{c2}(T)$ were fit. In addition, $H_{c2}(0)$ exceeded the Pauli limit for all three crystal axis directions, providing compelling evidence for parallel-spin state Cooper pairing.

    Upon the discovery of magnetic-field induced reentrant superconductivity in URhGe [18], much interest turned to the possible source of the high field superconducting phase. Then, superconductivity was discovered in UCoGe [12], and $H_{c2}(T)$ was measured for all three crystal axis directions [19], and all of the curves exhibited upward curvature [20]. Subsequently, a highly anomalous \textit{S}-shaped $H_{c2}(T)$ curve was observed for $T<0.65T_{c}$ with $\textbf{H}\|\textbf{b}$. Since $\textbf{M}\parallel \textbf{c}$ at low fields, this change in the $\textbf{M}$ direction only occurred in very pure, well aligned samples. This behavior may also have something to do with a reentrant phase, one that is close in field strength to the low-field phase [21,22].

    The first attempts to describe upward curvature in $H_{c2}(T)$ in all three crystal directions were based either upon ferromagnetic fluctuations [23], or upon a crossover from one parallel-spin state to another [24]. Meanwhile, a mean-field theory of the complementary effects of itinerant ferromagnetism and parallel-spin superconductivity was developed [25, 26]. To date, the field dependence of this mutual enhancement has not been investigated. Here, we study the case in which the \textit{p}-wave pairing interaction strength is anisotropic, but finite in all crystal directions. Since $H_{c2}(T)$ is essentially isotropic in the $ab$ plane for samples of UCoGe with medium purity [19], we studied the partially broken symmetry (PBS) state as a function of the pairing interaction anisotropy. This can give rise to a kink in $H_{c2}(T)$ in at least one crystal axis direction [27].
\section{Calculation}
Our calculation of $H_{c2}(T)$ assumes weak coupling for a homogeneous clean type-II material. $H_{c2}$ is therefore found by solving the linearized Gor'kov gap equation [2]

\begin{eqnarray}
\Delta(\mathbf{R},\hat{\mathbf{k}})&=&2\pi T\sum_{\omega_{n}}\int\frac{d\Omega_{k^{\prime}}}{4\pi}N(0)V(\hat{\mathbf{k}},\hat{\mathbf{k}}^{\prime})\int_{0}^{\infty}d\xi \nonumber\\
&&\times\textrm{exp}(-2\xi|\omega_{n}|)\textrm{exp}(-i\xi mv_{F}\textrm{sgn}\omega_{n}\hat{\mathbf{k}}^{\prime}\cdot\tensor{\bm M}\cdot\Pi)\nonumber\\
&&\times\Delta(\mathbf{R},\hat{\mathbf{k}}^{\prime}).
\end{eqnarray}
In our present case the interaction is of \textit{p}-wave form
\begin{equation}
V(\hat{\mathbf{k}},\hat{\mathbf{k}}^{\prime})=3\sum_{i}V_{i}\hat{\mathbf{k}}_{i}\cdot\hat{\mathbf{k}}_{i}^{\prime}|S_{i}(\hat{x}_{j})><S_{i}(\hat{x}_{j})|, \end{equation} which can be written in terms of the relevant spherical harmonics.
Here $N(0)$ is the single-spin density of states at the Fermi energy, $\Pi=\frac{\nabla}{i}-2e\mathbf{A}$, and
\begin{equation}
\tensor{\bm M}=\left(\begin{array}{ccc}
\sqrt{\frac{m}{m_{1}}} & 0 & 0\\
0 & \sqrt{\frac{m}{m_{2}}} & 0\\
0 & 0 & \sqrt{\frac{m}{m_{3}}}
\end{array}\right),
\end{equation}
where $m=(m_1m_2m_3)^{1/3}$ is the geometric mean effective mass.

We expand the order parameter in the harmonic oscillator states and the spherical harmonics
\begin{equation}
\Delta(\mathbf{R},\hat{\mathbf{k}})=\sum_{n=0}^{\infty}\sum_{m=-1}^{m=1}|n><n|\Delta_{1m}>Y_{1}^{m}(\theta_k,\phi_k),
\end{equation}
and insert Eq (4) into the Gor'kov gap equation. The details are presented in the appendix. Then for ${\bm B}\parallel\hat{\mathbf{e}_{3}}$, the polar and two axial PBS states are obtained from
\begin{equation}
<n|\Delta_{10}>\alpha_{n}^{(p)}=0,
\end{equation}
\begin{eqnarray}
\left(<n|\Delta_{11}>\pm<n|\Delta_{1,-1}>\right)\alpha_{n}^{\pm}&=&
\nonumber\\
\mp b_{n-2}<n-2|\Delta_{11}>-b_{n}<n+2|\Delta_{1,-1}>,&&
\end{eqnarray}
where 
\begin{eqnarray}
\alpha_{n}^{(p)}&=&\left[N(0)V_{3}\right]^{-1}-a_{n}^{(p)},\\  \alpha_{n}^{(-)}&=&\left[N(0)V_{2}\right]^{-1}-a_{n}^{(a)}, \\ \alpha_{n}^{(+)}&=&\left[N(0)V_{1}\right]^{-1}-a_{n}^{(a)},
\end{eqnarray}  and
\begin{eqnarray}
a_{n}^{(\lambda)}&=&
\pi T\sum_{\omega_{n}}\int_{0}^{\pi}d\theta \textrm{sin}\theta\left(\begin{array}{c}
3\textrm{cos} ^{2}\theta\\
\frac{3}{2}\textrm{sin}^{2}\theta
\end{array}\right)\int_{0}^{\infty}d\xi e^{-2\xi|\omega_{n}|}\nonumber \\
&&\times e^{-\frac{1}{2}\xi_{12}}
L_{n}(\xi_{12}),
\end{eqnarray}
\begin{equation}
b_{n}=\pi T\sum_{\omega_{n}}\int_{0}^{\pi}d\theta\frac{3}{2}\textrm{sin}^{2}\theta\int_{0}^{\infty}d\xi e^{-2\xi|\omega_{n}|}e^{-\frac{1}{2}\xi_{12}}F_{n}(\xi_{12}),
\end{equation}
where $\lambda=p$ $(a)$ corresponds to the upper (lower) row in Eq. (10), and $\xi_{12}=eB\xi^{2}v_{F}^{2}\mathrm{\textrm{sin}^{2}}\theta(\frac{m}{m_{12}})$, where $m_{12}=\sqrt{m_1m_2}$.  The bare transition temperatures are given by
\begin{eqnarray}
T_{ci}&=&\frac{2\gamma\omega_D}{\pi}e^{-1/N(0)V_i}
\end{eqnarray}
for $i=1,2,3$, where $\omega_D$ is the appropriate energy cutoff, and $\gamma\approx1.781$ is the exponential of Euler's constant.
Defining $x_{n}^{\pm}=<n|\Delta_{11}>\pm<n|\Delta_{1,-1}>$, we obtain two linear equations in the two variables $x_{n}^{\pm}$
\begin{equation}
2\alpha_{n}^{\pm}x_{n}^{\pm}\pm b_{n-2}(x_{n-2}^{+}+x_{n-2}^{-})+b_{n}(x_{n+2}^{+}-x_{n+2}^{-})=0,
\end{equation}
where $\alpha_{n}^{+}=\alpha_{n}^{-}+\delta$ and $\delta=\mathrm{ln}(\frac{T_{c1}}{T_{c2}})$, where $L_{n}(z)$ are the Laguerre polynomials, and $F_{n}(z)=\sum_{p=0}^{n}\frac{(-z)^{p+1}\sqrt{(n+2)(n+1)}n!}{p!(p+2)!(n-p)!}$.  With some algebraic manipulation shown in the appendix , one can show that these equations become a recursion relation
\begin{equation}
0=a_{n,n+2}x_{n+2}^{-}+a_{n,n}x_{n}^{-}+a_{n,n-2}x_{n-2}^{-},
\end{equation}
where
\begin{equation}
a_{n,n+2}=-\frac{\delta b_{n}}{\alpha_{n+2}^{+}}(1-c_{n-2}),
\end{equation}
\begin{equation}
a_{n,n-2}=-\frac{\delta b_{n-2}}{\alpha_{n-2}^{+}}(1-c_{n}),
\end{equation}
\begin{equation}
a_{n,n}=2\alpha_{n}^{-}(1-c_{n})(1-c_{n-2})-\delta(c_{n}+c_{n-2}-2c_{n}c_{n-2}),
\end{equation}
and
\begin{equation}
c_{n}=\frac{b_{n}^{2}}{\alpha_{n}^{+}\alpha_{n+2}^{+}-b_{n}^{2}}.
\end{equation}
It is shown in the appendix  that solving this recursion relation equation results in a continued fraction expression
\begin{equation}
a_{00}-\frac{a_{02}a_{20}}{a_{22}-\frac{a_{24}a_{42}}{a_{44\ldots}}}=0.
\end{equation}
By solving this equation for a given value of $\delta$, one obtains $h_{c2}$ for various PBS states ($\delta>0$, $\delta<0$) the ABM state ($\delta=0$), and the CBS state ($\delta=\infty$).
Since the low-field $H_{c2}(T)$ data of Huy \textit{et al.} for UCoGe suggests that it has uniaxial symmetry, we have restricted our study to cases of uniaxial anisotropy given $V_{1}=V_{2}\neq V_{3}$. A full orthorhombic symmetry treatment, in which $V_{1}\neq V_{2}\neq V_{3}$, would be necessary to fit the more anomalous \textit{S}-shaped $H_{c2}(T)$ observed by Aoki \textit{et al.} at higher externally applied fields. This will be treated in a future calculation, along with the inclusion of the spontaneous and field-induced magnetization. However, as a first approximation, we shall neglect the effects of the magnetization, as in the case of URhGe, in which the magnetization affects $B_{c2}(T)=\mu_0H_{c2}(T)$ by introducing a sharp demagnetization effect near $T_{c}$. 

In Fig. 1(a), we plotted the reduced field $h_{c2,\parallel c}(t)=2eB_{c2}(m/m_{12})v_{F}^{2}/(2\pi T_{c}^{c})^{2}$ versus the reduced temperature $t=T/T_{c}$  for the polar state and for a variety of PBSs with $-0.25\leq\delta<0$, where $\delta=\textrm{ln}(\frac{T_{c}^{ab}}{T_{c}^{c}})$. Note that these PBS
states all have an inherent upward curvature, but since $T_{c}^{c}>T_{c}^{ab}$, the polar state dominates near $T_{c}^{c}$. However, for $-0.05<\delta<0$ there is a single kink in $h_{c2,\parallel c}(t)$, and for $\delta=-0.0075$, there are two kinks, due to two crossovers between the polar and PBS states. For $\delta=-0.1$ there is no crossover to a PBS state. In Fig. 1(b), we plotted $h_{c2,\perp c}(t)=2eB_{c2}(m/\sqrt{m_{12}m_{3}})v_{F}^{2}/(2\pi T_{c}^{c})^{2}$ versus $t=T/T_{c}$ for the CBS state and for various PBSs with $-0.5\leq\delta<0$. In this case, the CBS state dominates near to $T_{c}^{c}$, but there is a crossover to the PBS state for $-0.3<\delta<0$, resulting in a single kink in $h_{c2,\perp c}(t)$. Below we discuss how these kinks result in upward curvature in all three crystal axis directions, and fits to the UCoGe data by Huy \textit{et al.} will be shown.
\begin{figure}
\center{\includegraphics[width=0.48\textwidth]{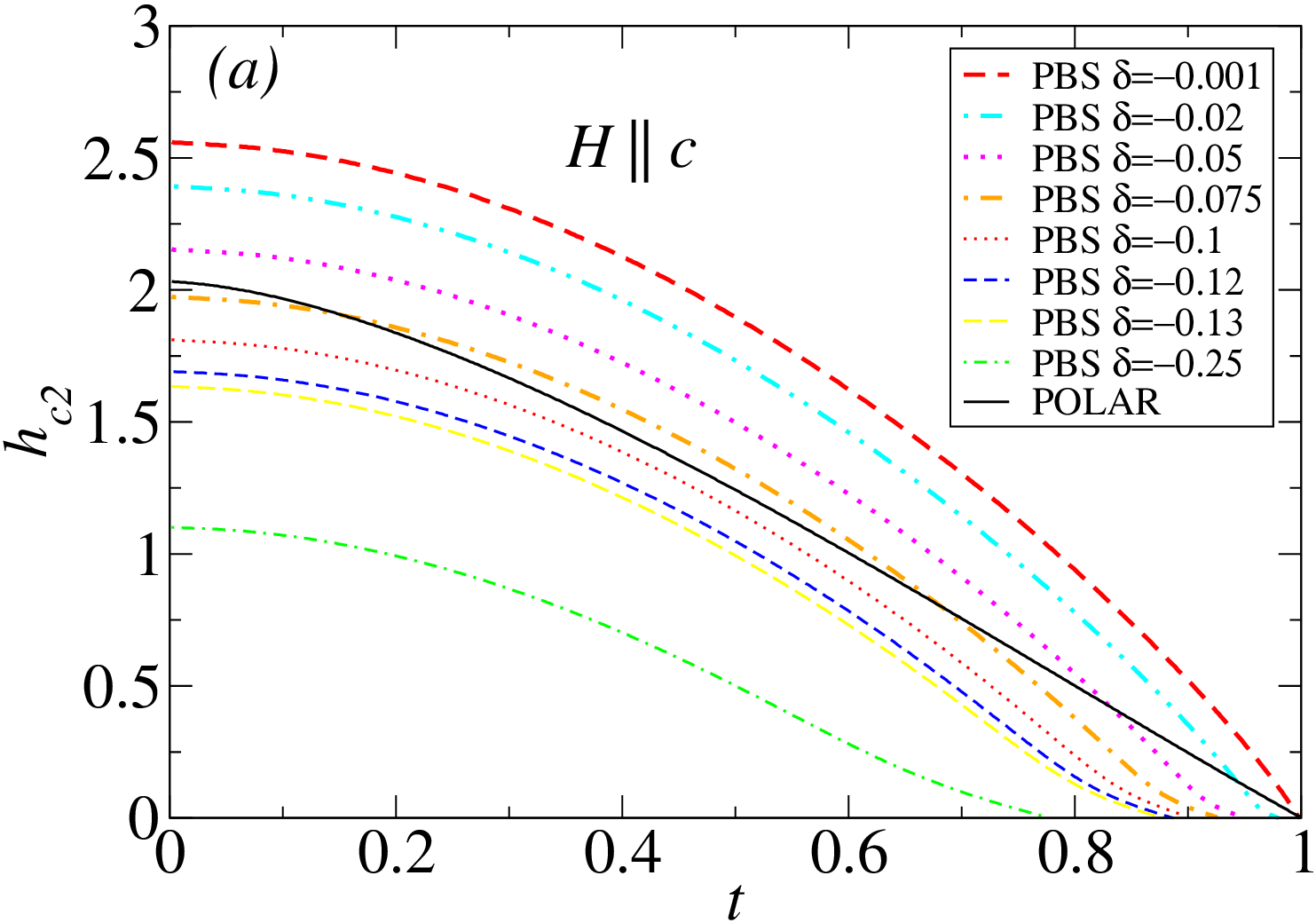}\vskip30pt
\includegraphics[width=0.48\textwidth]{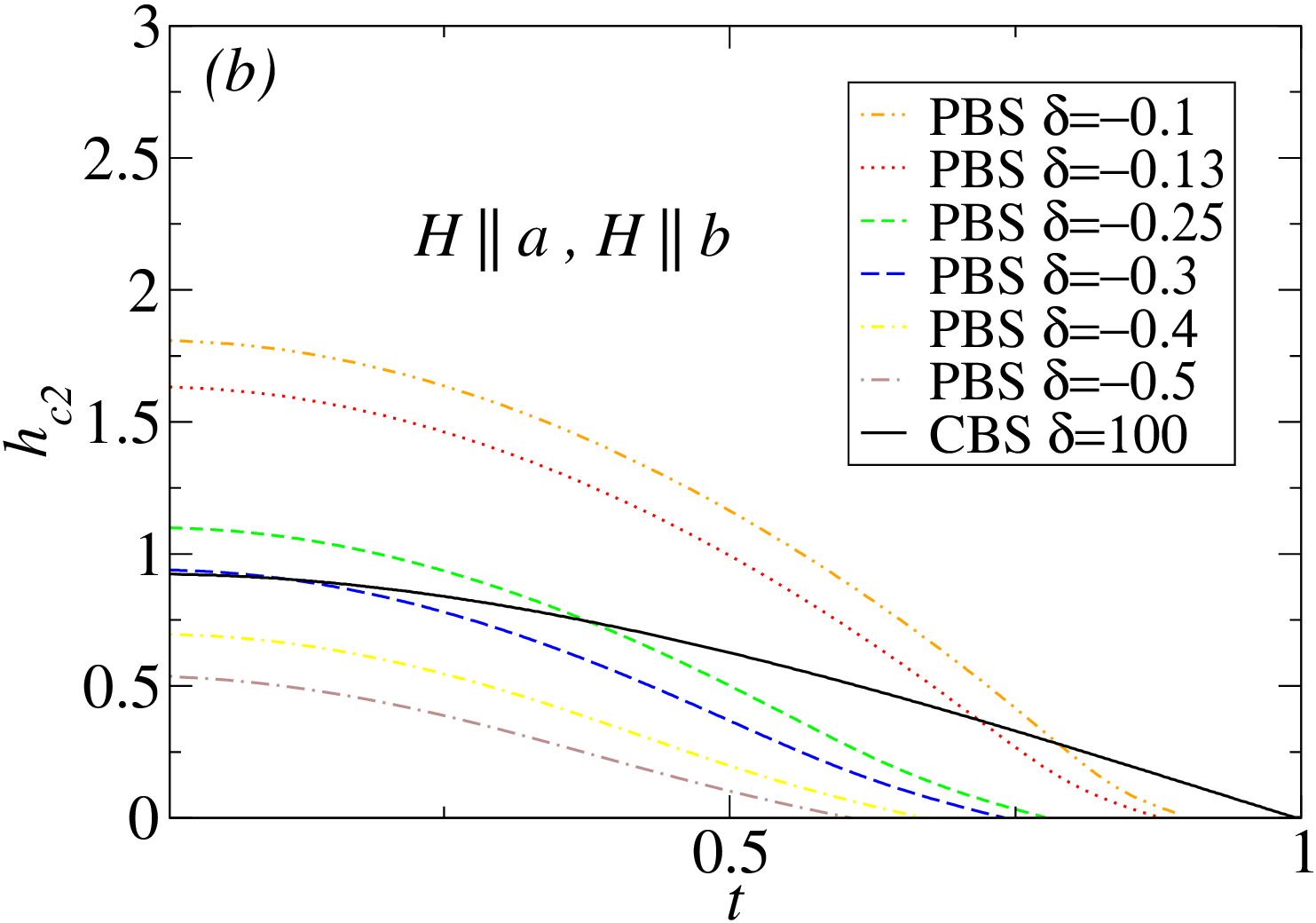}
\caption{(a) Plots of $h_{c2,||c}=2eB_{c2}(m/m_{12})v_F^2/(2\pi T^c_c)^2$ versus $t=T/T_c^c$ for the polar state (solid black) and for a variety of PBSs with $-0.25\le\delta=\ln(T_c^{ab}/T_c^c)\le-0.001$.  (b) Plots of $h_{c2,\perp c}=2eB_{c2}(m/\sqrt{m_{12}m_3})v_F^2/(2\pi T^c_c)^2$ versus $t=T/T_c^c$ for the CBS state (solid black) and for various PBSs with $-0.5\le\delta\le-0.1$.}}\label{fig1}
\end{figure}
\section{Fits to experimental data}
Using our uniaxial anisotropic model, we fit the least anomalous region of the upper critical field curves, where the symmetry in the $ab$ plane has not yet been broken by the field, obtained for medium purity samples of UCoGe using the data of Huy \textit{et al.} We have not included the spontaneous or field-induced magnetization into our calculations at this point. Future attempts to include the magnetization may prove to be fruitful, at the very least to show that it's effect on $H_{c2}(T)$ is minimal, as in URhGe.
\begin{figure}
\center{\includegraphics[width=0.4\textwidth]{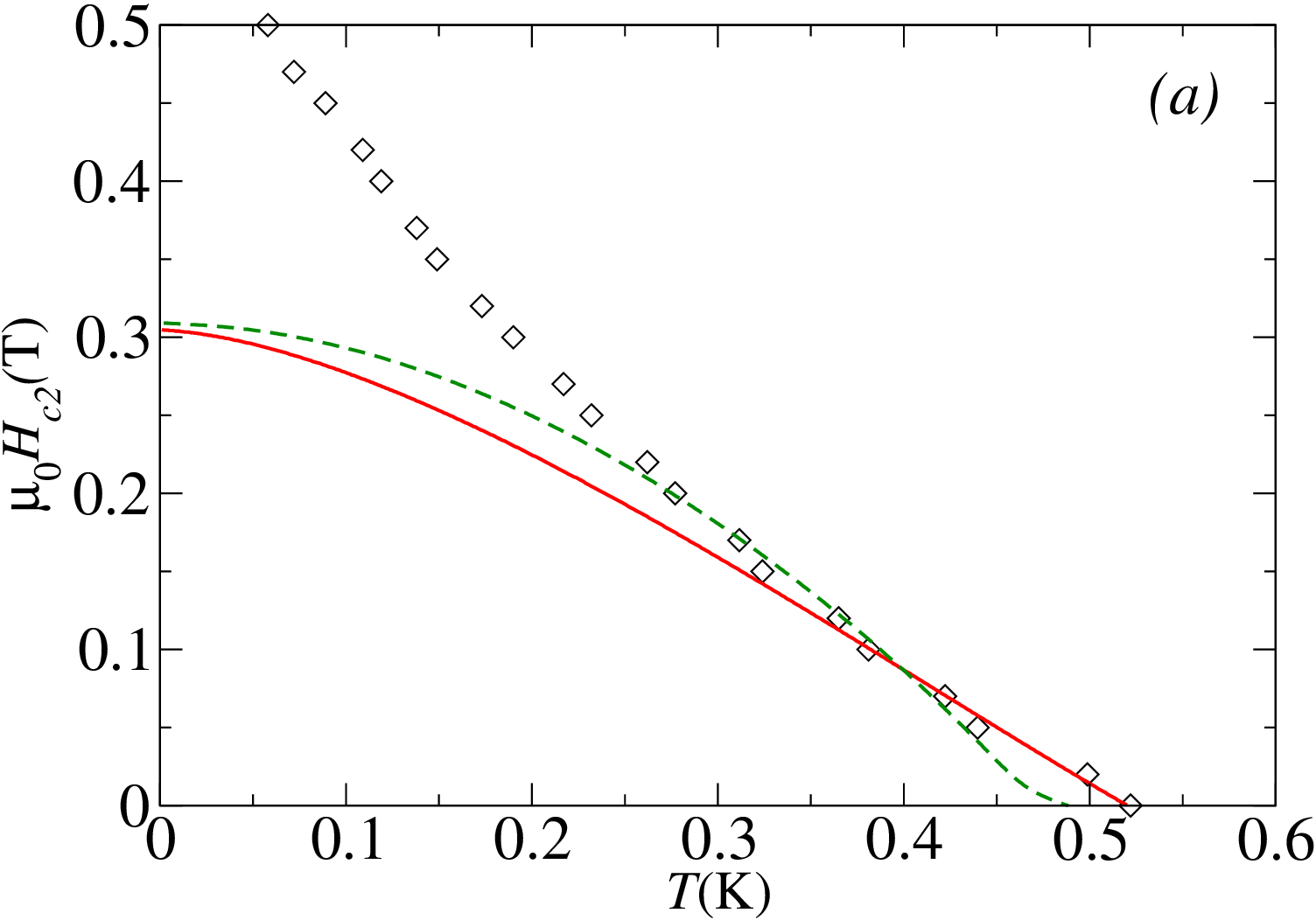}\vskip30pt
\includegraphics[width=0.4\textwidth]{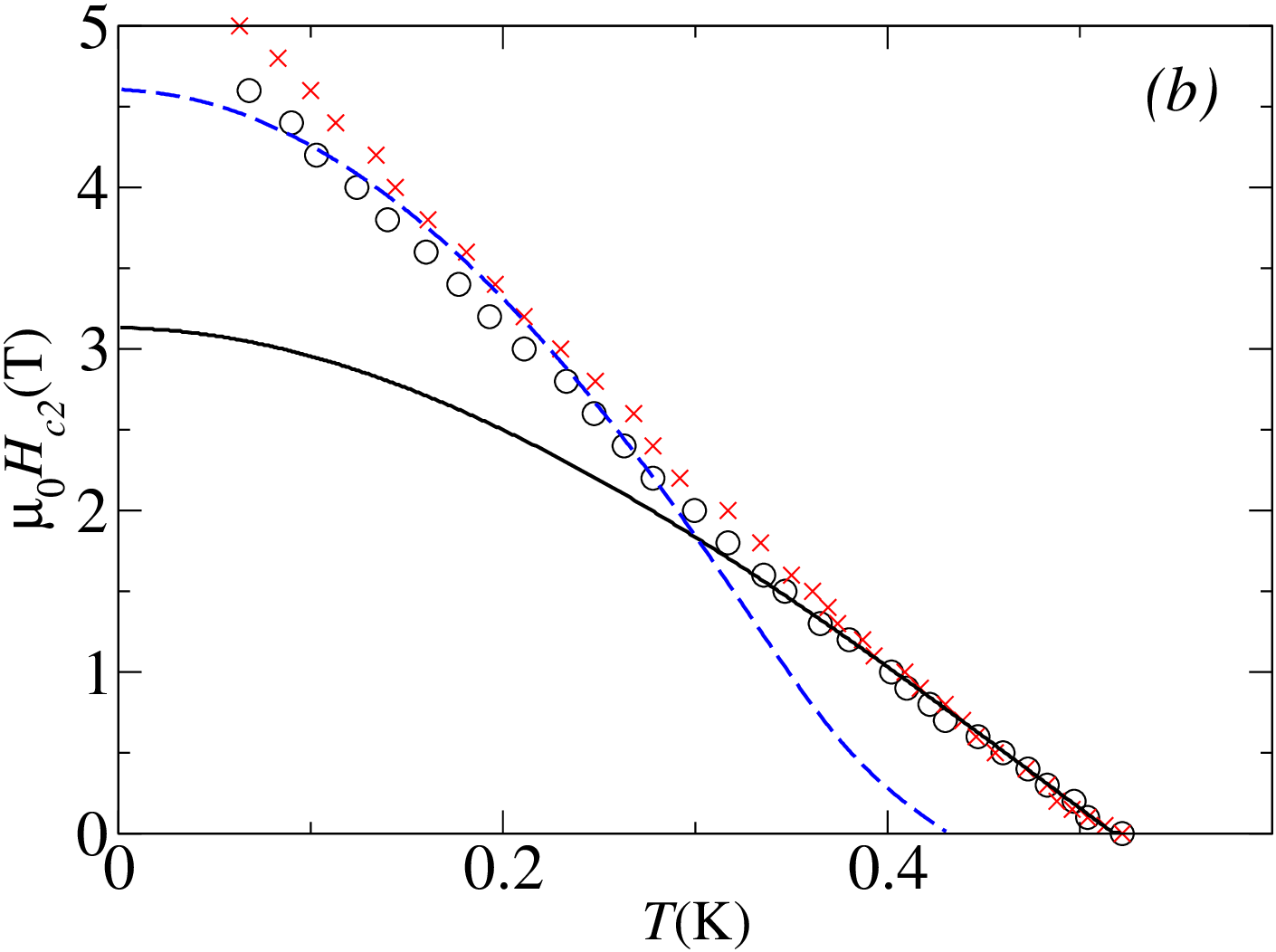}\caption{Best fits to the data  of Huy {\it et al.} for $\mu_0H_{c2}(T)$ in medium purity UCoGe [19]. (a) ${\bm H}||\hat{\bm c}$.  Open black diamonds: data. The red solid and green dashed curves are for the polar state and the PBS states with $\delta=-0.065$, respectively.  (b) Data for ${\bm H}||\hat{\bm b}$ (red crosses) and ${\bm H}||\hat{\bm a}$ (open black circles).  The solid black and blue dashed curves are for the CBS state and the PBS state with $\delta=-0.185$, respectively.  The slopes at $T_c$ were adjusted to fit the data.}}\label{fig2}
\end{figure}
In Fig. 2(a), the best fit to the data for the field parallel to the \textit{c}-axis is shown. The graph shows a distinct crossover from polar (solid red curve) to a PBS state with a $\delta=-0.0065$. The experimental data diverges from the theoretical values for larger fields. The inclusion of the spontaneous magnetization might alter the fit, but a more plausible explanation for this discrepancy is a field-dependent hopping term related to the effective electron mass. As is discussed in Aoki \textit{et al.}, this increase in effective mass may explain the \textit{S}-shaped curvature of UCoGe's upper critical field in the anomalous high field regime [21]. Proper consideration of these effects will be studied theoretically in the future. In Fig. 2(b), the best fit to the data for $\delta=-0.185$, which shows a distinct field-induced crossover from the CBS to the PBS state.

Using extremely clean single crystal samples, Aoki \textit{et al.} has shown that the field alignment affects the upward curvature in UCoGe quite substantially, especially at high fields, which is likely due to the field-induced ferrimagnetic state. It is therefore of great importance to calculate the angle dependence of the slope of the upper critical field, at least close to $T_{c}$. Though the angle dependence of the polar state was analytically shown to be proportional to $[\cos^{2}\theta+\frac{3m_{ab}}{m_{c}}\sin^{2}\theta]^{-1/2}$ by Scharnberg and Klemm [2], the angle dependence of the axial state has not yet been calculated analytically. This angle dependence may prove useful in identifying the $p$-wave superconducting state present in certain materials, and possibly suggest new experiments on the well known \textit{p}-wave superconductor URhGe.

\section{Conclusions}
We found that it is possible to fit the upward curvature
of the $H_{c2}(T)$ data from medium-purity UCoGe using a
crossover from the polar/CBS state to a PBS state. However,
without taking account of the spontaneous magnetization, the fitting parameter $\delta$ was different for the field in the $c$-axis direction than it was for the field in either the $a$ or $b$ directions.
 At the very
least, the spontaneous and field-dependent magnetization
should be included in future fits, using an anisotropic itinerant
ferromagnetic superconductor model similar to
that previously studied.
\section{Appendix}

We start with the linearized Gor'kov gap equation, Eq. (1), and 
and expand $\Delta(\mathbf{R},\hat{\mathbf{k}})$ in the spherical harmonics and the harmonic oscillator basis states as in Eq. (4).
By projecting out the $\Delta_{10}$ component one obtains

\begin{eqnarray}
<n^{\prime}|\Delta_{10}>&=&2\pi TN(0)V_{3}\frac{3}{2}\sum_{\omega_{n}}\int d\theta_{k^{\prime}}\textrm{sin}\theta_{k^{\prime}}\textrm{cos}^{2}\theta_{k^{\prime}}\nonumber\\
&&\int_{0}^{\infty}d\xi e^{-2\xi|\omega_{n}|}e^{-\frac{1}{2}eH\xi^{2}v_{F}^{2}\textrm{sin}^{2}\theta_{k^{\prime}}(\frac{m}{m_{12}})}\nonumber \\
&&\times <n^{\prime}|e^{-\alpha^{*}a^{\dagger}}e^{\alpha a}|n^{\prime}><n^{\prime}|\Delta_{10}>,
\end{eqnarray}
where $\alpha=i\xi v_{F}\sqrt{\frac{eHm}{m_{12}}}\textrm{sin}\theta_{k^{\prime}}e^{i\phi_{k^{\prime}}}$ and we have used the well known identity $e^{\hat{a}+\hat{b}}=e^{\hat{a}}e^{\hat{b}}e^{-\frac{1}{2}\left[\hat{a},\hat{b}\right]}$, which is valid for $\left[\hat{a},\left[\hat{a},\hat{b}\right]\right]=\left[\hat{b},\left[\hat{a},\hat{b}\right]\right]=0$.
Similarly, one can find expressions for the other two states,

\begin{eqnarray}
<n^{\prime}|\Delta_{11}>&=&2\pi TN(0)\frac{3}{4}V_{2}\sum_{\omega_{n}}\int_{0}^{\pi}d\theta_{k^{\prime}}\textrm{sin}^{3}\theta_{k^{\prime}} \nonumber\\
-<n^{\prime}|\Delta_{1,-1}>&&\times \int_{0}^{2\pi}\frac{d\phi_{k^{\prime}}}{2\pi}\int_{0}^{\infty}d\xi e^{-2\xi|\omega_{n}|}\nonumber\\ &&\times e^{-\frac{1}{2}eH\xi^{2}v_{F}^{2}\textrm{sin}^{2}\theta_{k^{\prime}}\frac{m}{m_{12}}}\nonumber\\
& &\times\Bigl[<n^{\prime}|e^{-\alpha^{*}a^{\dagger}}e^{\alpha a}|n^{\prime}>\nonumber\\
& &\times\Bigl(<n^{\prime}|\Delta_{11}>-<n^{\prime}|\Delta_{1,-1}>\Bigr) \nonumber \\
&&-<n^{\prime}|e^{-\alpha^{*}a^{\dagger}}e^{\alpha a}|n^{\prime}+2>\nonumber\\
& &\times<n^{\prime}+2|\Delta_{1,-1}>e^{-2i\phi_{k^{\prime}}}\nonumber\\
& &+<n^{\prime}|e^{-\alpha^{*}a^{\dagger}}e^{\alpha a}|n^{\prime}-2>\nonumber\\
& &\times<n^{\prime}-2|\Delta_{11}>e^{2i\phi_{k^{\prime}}}\Bigr]
\end{eqnarray}

\begin{eqnarray}
<n^{\prime}|\Delta_{11}>&=&2\pi TN(0)\frac{3}{4}V_{1}\sum_{\omega_{n}}\int_{0}^{\pi}d\theta_{k^{\prime}}\textrm{sin}^{3}\theta_{k^{\prime}} \nonumber\\
+<n^{\prime}|\Delta_{1,-1}>&&\times \int_{0}^{2\pi}\frac{d\phi_{k^{\prime}}}{2\pi}\int_{0}^{\infty}d\xi e^{-2\xi|\omega_{n}|}\nonumber\\ &&\times e^{-\frac{1}{2}eH\xi^{2}v_{F}^{2}\textrm{sin}^{2}\theta_{k^{\prime}}\frac{m}{m_{12}}}\nonumber\\
& &\times\Bigl[<n^{\prime}|e^{-\alpha^{*}a^{\dagger}}e^{\alpha a}|n^{\prime}>\nonumber\\
& &\times\Bigl(<n^{\prime}|\Delta_{11}>+<n^{\prime}|\Delta_{1,-1}>\Bigr) \nonumber \\
&&-<n^{\prime}|e^{-\alpha^{*}a^{\dagger}}e^{\alpha a}|n^{\prime}+2>\nonumber\\
& &\times<n^{\prime}+2|\Delta_{1,-1}>e^{-2i\phi_{k^{\prime}}}\nonumber\\
& &-<n^{\prime}|e^{-\alpha^{*}a^{\dagger}}e^{\alpha a}|n^{\prime}-2>\nonumber\\
& &\times<n^{\prime}-2|\Delta_{11}>e^{2i\phi_{k^{\prime}}}\Bigr]
\end{eqnarray}

It can be shown that $<n^{\prime}|e^{-\alpha^{*}a^{\dagger}}e^{\alpha a}|n^{\prime}>=L_{n^{\prime}}(|\alpha|^{2})$ and $<n^{\prime}|e^{-\alpha^{*}a^{\dagger}}e^{\alpha a}|n^{\prime}-2>=e^{-2i\phi_{k^{\prime}}}F_{n^{\prime}-2}(|\alpha|^{2})$ and $<n^{\prime}|e^{-\alpha^{*}a^{\dagger}}e^{\alpha a}|n^{\prime}+2>=e^{2i\phi_{k^{\prime}}}F_{n^{\prime}}(|\alpha|^{2})$ where $L_{n}(x)$ are the Laguerre Polynomials and $F_{n}(z)=\sum_{p=0}^{n}\frac{(-z)^{p+1}\sqrt{(n+2)(n+1)}n!}{p!(p+2)!(n-p)!}$ from which we can derive the equations for $H_{c2}(T)$. Equations (20)-(22) become

\begin{eqnarray}
<n^{\prime}|\Delta_{10}>&=&2\pi TN(0)V_{3}\frac{3}{2}\sum_{\omega_{n}}\int_{0}^{\pi}d\theta_{k^{\prime}}\textrm{sin}\theta_{k^{\prime}}\textrm{cos}^{2}\theta_{k^{\prime}}
\nonumber \\
&&\times \int_{0}^{\infty}d\xi  e^{-2\xi|\omega_{n}|}e^{-\frac{1}{2}eH\xi^{2}v_{F}^{2}\textrm{sin}^{2}\theta_{k^{\prime}}(\frac{m}{m_{12}})} \nonumber\\
&&\times L_{n^{\prime}}(|\alpha|^{2})
<n^{\prime}|\Delta_{10}>
\end{eqnarray}

\begin{eqnarray}
<n^{\prime}|\Delta_{11}>&=&2\pi TN(0)\frac{3}{4}V_{2}\sum_{\omega_{n}}\int_{0}^{\pi}d\theta_{k^{\prime}}\textrm{sin}^{3}\theta_{k^{\prime}} \nonumber\\
-<n^{\prime}|\Delta_{1,-1}>&&\times \int_{0}^{2\pi}\frac{d\phi_{k^{\prime}}}{2\pi}\int_{0}^{\infty}d\xi e^{-2\xi|\omega_{n}|}\nonumber\\ &&\times e^{-\frac{1}{2}eH\xi^{2}v_{F}^{2}\textrm{sin}^{2}\theta_{k^{\prime}}\frac{m}{m_{12}}}\nonumber\\
& &\times\Bigl[L_{n^{\prime}}(|\alpha|^{2})\nonumber\\
& &\times\Bigl(<n^{\prime}|\Delta_{11}>-<n^{\prime}|\Delta_{1,-1}>\Bigr) \nonumber \\
&&-F_{n^{\prime}}(|\alpha|^{2})<n^{\prime}+2|\Delta_{1,-1}>\nonumber\\
& &+F_{n^{\prime}-2}(|\alpha|^{2})
<n^{\prime}-2|\Delta_{11}>\Bigr]
\end{eqnarray}

\begin{eqnarray}
<n^{\prime}|\Delta_{11}>&=&2\pi TN(0)\frac{3}{4}V_{1}\sum_{\omega_{n}}\int_{0}^{\pi}d\theta_{k^{\prime}}\textrm{sin}^{3}\theta_{k^{\prime}} \nonumber\\
+<n^{\prime}|\Delta_{1,-1}>&&\times \int_{0}^{2\pi}\frac{d\phi_{k^{\prime}}}{2\pi}\int_{0}^{\infty}d\xi e^{-2\xi|\omega_{n}|}\nonumber\\ &&\times e^{-\frac{1}{2}eH\xi^{2}v_{F}^{2}\textrm{sin}^{2}\theta_{k^{\prime}}\frac{m}{m_{12}}}\nonumber\\
& &\times\Bigl[L_{n^{\prime}}(|\alpha|^{2})\nonumber\\
& &\times\Bigl(<n^{\prime}|\Delta_{11}>+<n^{\prime}|\Delta_{1,-1}>\Bigr) \nonumber \\
&&-F_{n^{\prime}}(|\alpha|^{2})<n^{\prime}+2|\Delta_{1,-1}>\nonumber\\
& &-F_{n^{\prime}-2}(|\alpha|^{2})
<n^{\prime}-2|\Delta_{11}>\Bigr]
\end{eqnarray}

These equations may be rewritten as
\begin{equation}
<n|\Delta_{10}>\alpha_{n,12}^{(p)}=0,
\end{equation}
\begin{equation}
2\alpha_{n}^{\pm}x_{n}^{\pm}\pm b_{n-2}(x_{n-2}^{+}+x_{n-2}^{-})+b_{n}(x_{n+2}^{+}-x_{n+2}^{-})=0,
\end{equation}
where $x_{n}^{\pm}$ was defined following Eq. (12) in the text.

We note that the expression for the polar state separates from that for the two axial states, which comprise the PBS state for $V_1\ne V_2$. 
In order to obtain the recursion relations for the PBS state, we begin by 
adding and subtracting both equations in Eq. (13).  This leads to

\begin{equation}
\alpha_{n}^{+}x_{n}^{+}+b_{n}x_{n+2}^{+}=b_{n}x_{n+2}^{-}-\alpha_{n}^{-}x_{n}^{-}
\end{equation}
\begin{equation}
\alpha_{n}^{+}x_{n}^{+}+b_{n-2}x_{n-2}^{+}=\alpha_{n}^{-}x_{n}^{-}-b_{n-2}x_{n-2}^{-},
\end{equation}

where we can let $n\rightarrow n-2$  in Eq (28) to obtain
\begin{equation}
b_{n-2}x_{n}^{+}+\alpha_{n-2}^{+}x_{n-2}^{+}=b_{n-2}x_{n}^{-}-\alpha_{n-2}^{-}x_{n-2}^{-}
\end{equation}
\begin{equation}
\alpha_{n}^{+}x_{n}^{+}+b_{n-2}x_{n-2}^{+}=\alpha_{n}^{-}x_{n}^{-}-b_{n-2}x_{n-2}^{-}
\end{equation}

We now have two equations in the two unknowns $x_{n}^{+}$ and $x_{n-2}^{+}$, resulting in

\begin{equation}
x_{n}^{+}=x_{n}^{-}\left(1-\frac{\delta\alpha_{n-2}^{+}}{\alpha_{n}^{+}\alpha_{n-2}^{+}-b_{n-2}^{2}}\right)-\frac{\delta b_{n-2}x_{n-2}^{-}}{\alpha_{n}^{+}\alpha_{n-2}^{+}-b_{n-2}^{2}}
\end{equation}
\begin{equation}
x_{n-2}^{+}=x_{n-2}^{-}\left(-1+\frac{\delta\alpha_{n}^{+}}{\alpha_{n}^{+}\alpha_{n-2}^{+}-b_{n-2}^{2}}\right)+\frac{\delta b_{n-2}x_{n}^{-}}{\alpha_{n}^{+}\alpha_{n-2}^{+}-b_{n-2}^{2}}
\end{equation}

Letting $n\rightarrow n+2$ in Eq. (33) and setting the two equations equal, we obtain
\begin{eqnarray}
0&=&x_{n}^{-}\left(-2+\frac{\delta\alpha_{n+2}^{+}}{\alpha_{n+2}^{+}\alpha_{n}^{+}-b_{n}^{2}}
+\frac{\delta\alpha_{n-2}^{+}}{\alpha_{n}^{+}\alpha_{n-2}^{+}-b_{n-2}^{2}}\right)\nonumber\\
&&+\frac{\delta b_{n}x_{n+2}^{-}}{\alpha_{n+2}^{+}\alpha_{n}^{+}-b_{n}^{2}}+\frac{\delta b_{n-2}x_{n-2}^{-}}{\alpha_{n}^{+}\alpha_{n-2}^{+}-b_{n-2}^{2}},
\end{eqnarray}
where $\delta=\ln(T_{c1}/T_{c2})$, 
from which we obtain the recursion relation

\begin{equation}
0=a_{n,n+2}x_{n+2}^{-}+a_{n,n}x_{n}^{-}+a_{n,n-2}x_{n-2}^{-},
\end{equation}
where

\begin{equation}
a_{n,n+2}=-\frac{\delta b_{n}}{\alpha_{n+2}^{+}}(1-c_{n-2}),
\end{equation}
\begin{equation}
a_{n,n-2}=-\frac{\delta b_{n-2}}{\alpha_{n-2}^{+}}(1-c_{n}),
\end{equation}
\begin{equation}
a_{n,n}=2\alpha_{n}^{-}(1-c_{n})(1-c_{n-2})-\delta(c_{n}+c_{n-2}-2c_{n}c_{n-2}),
\end{equation}
and
\begin{equation}
c_{n}=\frac{b_{n}^{2}}{\alpha_{n}^{+}\alpha_{n+2}^{+}-b_{n}^{2}}.
\end{equation}
We note that there was an unfortunate typo in the expression for $c_n$ in Ref. (2).

The solution to the recursion relation can be shown to be a continued fraction equation involving the reduced field $h$ and the reduced temperature $t$

\begin{equation}
a_{00}-\frac{a_{02}a_{20}}{a_{22}-\frac{a_{24}a_{42}}{a_{44\ldots}}}=0.
\end{equation}

Solutions to this equation are obtained using standard numerical techniques. It is important to note that the first order iteration differs from the exact curve by less than $2\%$. It is therefore only necessary to consider the continued fraction up to the $a_{22}$ term, as the iteration thereafter converges very rapidly.

We note that a very brief version of this work appeared previously [28].

\section{Acknowledgments}
\begin{acknowledgments}
The authors are grateful to Prof. A. de Visser for providing the data of Huy et al. CL acknowledges the Florida Education Fund and the McKnight Doctoral Fellowship for their support. QG acknowledges the Specialized Research Fund for the Doctoral Program of Higher Education of China (no. 20100006110021).
\end{acknowledgments}

\subsection{}
\subsubsection{}

\end{document}